\documentclass{elsart}

\usepackage{epsfig}

\usepackage{amssymb}

\begin{document}

\begin{frontmatter}


\title{Measurement of $\Gamma(K_{\mu 3})/\Gamma(K_{e3})$ ratio using stopped
positive kaons }

\thanks[email]{ Corresponding author.\\{\it E-mail address:}
 kate@phys.wani.osaka-u.ac.jp~(Keito HORIE)}

\author[osa]{K.~Horie \thanksref{email}},
\author[osa]{S.~Shimizu},
\author[tuku1]{M.~Abe}, 
\author[kek]{M.~Aoki}, 
\author[tuku2]{I.~Arai}, 
\author[tuku1]{Y.~Asano}, 
\author[kek]{T.~Baker},
\author[mar]{M.~Blecher}, 
\author[kek]{M.D.~Chapman},
\author[dep]{P.~Depommier}, 
\author[has]{M.~Hasinoff}, 
\author[tai]{H.C.~Huang}, 
\author[tuku2,kek]{Y.~Igarashi \thanksref{sugu}},
\author[tuku2,rik]{T.~Ikeda \thanksref{sugu}},
\author[kek]{J.~Imazato}, 
\author[inr]{A.P.~Ivashkin}, 
\author[inr]{M.M.~Khabibullin}, 
\author[inr]{A.N.~Khotjantsev},
\author[inr]{Y.G.~Kudenko},
\author[kek,osa]{Y.~Kuno \thanksref{sugu}},
\author[yon]{J.-M.~Lee}, 
\author[kor]{K.S.~Lee}, 
\author[inr]{A.S. Levchenko},
\author[kek]{G.Y.~Lim}, 
\author[tri]{J.A.~Macdonald}, 
\author[pri]{C.R.~Mindas},
\author[inr]{O.V.~Mineev}, 
\author[yon]{Y.-H.~Shin}, 
\author[tri]{Y.-M.~Shin}, 
\author[tuku2]{A.~Suzuki}, 
\author[tuku2]{A.~Watanabe}, 
 and 
\author[yoko,kek]{T.~Yokoi \thanksref{sugu}}

\collab{KEK-E246 Collaboration}
\address[osa]{Department of Physics, Osaka University, Osaka 560-0043,
 Japan }
\address[tuku1]{
 Institute of Applied Physics, University of Tsukuba, Ibaraki 305-0006, 
 Japan }
\address[kek]{
 IPNS, High Energy Accelerator Research Organization (KEK), Ibaraki 
 305-0801, Japan }
\address[tuku2]{
 Institute of Physics, University of Tsukuba, Ibaraki 305-0006, Japan }
\address[mar]{
 Department of Physics, Virginia Polytechnic Institute and State University, 
      VA 24061-0435, U.S.A. }
\address[dep]{
Laboratoire de Physique Nucl\'eaire, Universit\'e de Montr\'{e}al,
Montr\'{e}al, Qu\'ebec, Canada H3C 3J7}
\address[inr]{
Institute for Nuclear Research,
Russian Academy of Sciences, Moscow 117312, 
      Russia }
\address[has]{
 Department of Physics and Astronomy, University of British Columbia, 
 Vancouver, Canada V6T 1Z1 }
\address[tai]{Department of Physics, National Taiwan University, Taipei
 106, Taiwan}
\address[rik]{The Institute of Physical and Chemical Research, Saitama
 351-0106, Japan.}
\address[yon]{ 
Department of Physics, Yonsei University, Seoul 120-749, Korea}
\address[kor]{
Department of Physics, Korea University, Seoul 136-701, Korea }
\address[tri]{
TRIUMF, Vancouver, British Columbia, Canada V6T 2A3 }
\address[pri]{
Department of Physics, Princeton University, NJ 08544, U.S.A.}
\address[yoko]{Department of Physics, University of Tokyo, Tokyo
 113-0033, Japan}

\thanks[sugu]{Second institutions are present addresses.}

 \begin{abstract}
  The ratio of the $K^{+}\rightarrow \pi^{0} \mu^{+} \nu$ ($K_{\mu3}^+$)
  and $K^{+}\rightarrow \pi^{0} e^{+} \nu$ ($K_{e3}^+$) decay 
  widths, $\Gamma(K_{\mu 3})/\Gamma(K_{e3})$, has been measured with
  stopped positive kaons. $K_{\mu3}^+$ and $K_{e3}^+$ samples containing
  2.4$\times 10^4$ and 4.0$\times 10^4$ events, respectively, were
  analyzed. The $\Gamma(K_{\mu3})/\Gamma(K_{e3})$ ratio was obtained to
  be 0.671$\pm$0.007(stat.)$\pm$0.008(syst.) calculating the detector
  acceptance by a Monte Carlo simulation.
  The coefficient of the
  $q^2$ dependent term of the $f_0$ form factor was also determined to
  be $\lambda_0$=0.019$\pm$0.005(stat.)$\pm$0.004(syst.)
  with the assumption of
  $\mu$-$e$ universality in $K_{l3}^+$ decay. 
  The agreement
  of our result with the $\lambda_0$ value obtained from $K_{\mu 3}^+$
  Dalitz plot analyses supports the validity of the $\mu$-$e$
  universality.  
 \end{abstract}


\end{frontmatter}
\section{Introduction}
The spectroscopic studies to determine form factors of the $K^+$
semi-leptonic decays, $K^{+}\rightarrow
\pi^{0} l^{+} \nu$ ($K_{l 3}^+$), are of importance both in studying 
low energy properties of the strong interaction in terms of effective
theories~\cite{gas85,kaz76}, and also in studying fundamental
interactions. In our previous work~\cite{shi00}, we reported a result
testing the exotic couplings in $K^{+}\rightarrow \pi^{0} e^{+} \nu$
($K_{e3}^+$) decay, showing the non existence of scalar and tensor
interactions, contradicting the current world average adopted by
Particle Data Group~\cite{par00}. In the present work, the
$K^{+}\rightarrow \pi^{0} \mu^{+} \nu$ ($K_{\mu3}^+$) events, which were
collected simultaneously, were analyzed to determine the ratio of the
$K_{\mu3}^+$ and $K_{e3}^+$ decay widths
$\Gamma(K_{\mu3})/\Gamma(K_{e3})$. This quantity is one of the most
important observables to evaluate the $K_{l3}^+$ form factors. 

Assuming that only the V$-$A interaction contributes to the $K_{l3}$
decay, the decay amplitude can be described by two dimensionless
form factors, $f_{+}(q^{2})$ and $f_{0}(q^{2})$, 
which are functions of the 
momentum transfered to the leptons  
$q^{2}$=$(P_{K}-P_{\pi^{0}})^2$ 
 where 
 $P_{K}$ and $P_{\pi^0}$ are the four momenta of the $K^{+}$ and $\pi^{0}$, 
respectively. They are given as, 
\begin{eqnarray}
f_{+}(q^{2})&=&f_{+}(0)[1+\lambda_{+}(q/m_{\pi})^{2}], \nonumber \\
f_{0}(q^{2})&=&f_{0}(0)[1+\lambda_{0}(q/m_{\pi})^{2}]. \nonumber
\end{eqnarray}
Assuming $\mu$-$e$ universality, the
form factors between $K_{\mu 3}$ and $K_{e3}$ decays
are identical and  
$\Gamma(K_{\mu3})/\Gamma(K_{e3})$ can be written as~\cite{fea70},
\begin{eqnarray}
\Gamma(K_{\mu 3})/\Gamma(K_{e3})&=&0.6457-0.1531\lambda_+ \nonumber\\
&&+1.5646\lambda_0 + O(\lambda_+^2+\lambda_+\lambda_0+\lambda_0^2). \label{kihon1}
\end{eqnarray}
This equation cannot determine the $\lambda_+$ and $\lambda_0$
parameters uniquely but simply fixes a relationship between
them. However, if the $\lambda_+$ value derived from $K_{e3}$ data
analyses is assumed, the $\lambda_0$ parameter can be obtained from
Eq.~(\ref{kihon1}). In this analysis, our $K_{e3}$ result for the
$\lambda_+$ parameter, $\lambda_+=0.0278\pm0.0040$, was
employed~\cite{shi00}. 

It should be noted that the $\lambda_0$ parameter can be also determined
by studying the Dalitz plot distribution of $K_{\mu 3}$ decay or the
muon polarization in $K_{\mu 3}$ decay~\cite{par00}. The determination of the
$\lambda_0$ parameter from the $\Gamma(K_{\mu 3})/\Gamma(K_{e3})$ ratio
is based on the assumption of $\mu$-$e$ universality. This is not the
case for the other two methods. Thus, $\mu$-$e$ universality in $K_{l3}$
decay can be tested by comparing the $\lambda_0$ parameter obtained from
a branching ratio measurement $\lambda_0^{\rm br}$ with that obtained
from a Dalitz plot measurement $\lambda_0^{\rm dal}$ and/or $\mu^+$
polarization measurement $\lambda_0^{\rm pol}$. 
In this case, however, a precise acceptance function has to
be carefully determined over the whole Dalitz space for the
$K_{\mu 3}$ Dalitz plot analysis.

In this letter, we present a new precise measurement of the
$\Gamma(K_{\mu 3})/\Gamma(K_{e3})$ branching ratio and the $\lambda_0$
parameter. The experiment used a stopped $K^+$ beam in conjunction with
a 12-sector iron-core superconducting toroidal
spectrometer~\cite{ima:90}. $\Gamma(K_{\mu 3})/\Gamma(K_{e3})$ was
obtained by
measuring the ratio of the number of accepted $K_{\mu3}$ and $K_{e3}$
events corrected for detector acceptance.  

\section{Experiment}
\begin{figure}
\begin{center}
\epsfig{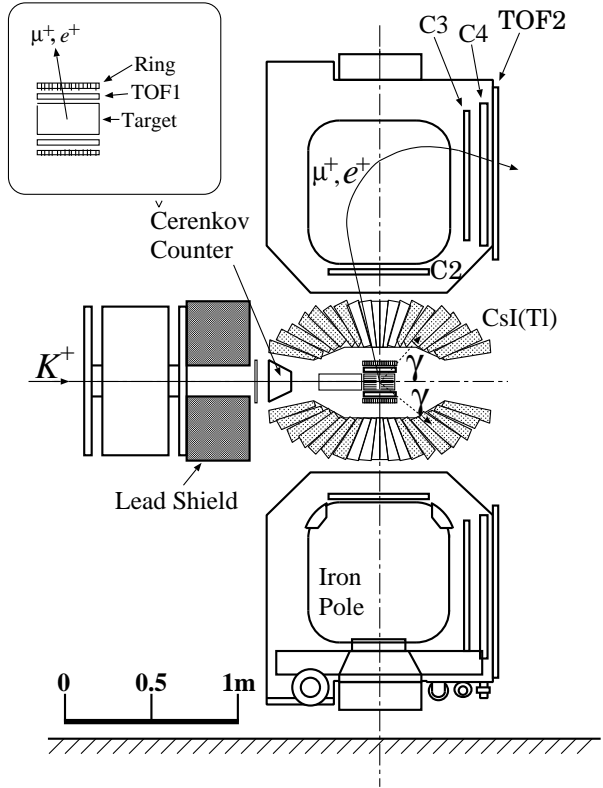}
\caption{ 
Cross sectional side view of the E246 setup. Assembly detail of the active
 target, TOF1, and ring counter is shown in the inset.
}
\label{fg:side}
\end{center}
\end{figure}

The experiment was performed at the KEK 12 GeV proton synchrotron.
The experimental apparatus was constructed for a
T-violation search in $K_{\mu3}$ decay~\cite{main:99}. 
A schematic cross sectional side view of the detector is shown in
Fig.~\ref{fg:side}.
We detected both
of the $\pi^0$ decay gammas which enabled us to reconstruct the complete
kinematics of a $K_{l3}$ event in contrast to the previous
experiment~\cite{heintze}; this increased the reliability of the event
selection. 
Because of the rotational symmetry of the 12 identical gaps in the
spectrometer and the large directional acceptance of the $\pi^0$
detector~\cite{nimcsi}, distortions due to detector acceptance were
drastically reduced. Moreover, the similarity of the $K_{\mu 3}$ and
$K_{e 3}$ kinematics reduced the systematic error due to the imperfect
reproducibility of the experimental conditions in the simulation of the
ratio of the calculated acceptance for the $K_{\mu 3}$ and $K_{e 3}$
decays. The measurement was carried out for two values of the central
magnetic field strength, B=0.65 and 0.90 T, yielding a consistency check
with regard to the spectrometer acceptance and energy loss estimation in
the target.  

$K_{\mu 3}$ ($K_{e 3}$) events were identified by analyzing the $\mu^+$
($e^+$) momentum with the spectrometer and detecting the two photons in
the CsI(Tl) calorimeter. Charged particles from the target were tracked
and momentum-analyzed using multi-wire proportional chambers (MWPCs) at
the entrance (C2) and exit (C3 and C4) of the magnet gap, as well as by
the active target and an array of ring counters~\cite{ring:97}
surrounding the target. Particle identification between the $\mu^{+}$s
and $e^+$s was carried out by time-of-flight (TOF) between TOF1 and TOF2
scintillation counters. TOF1 surrounds the active target and TOF2 is
located at the exit of the spectrometer. The $\pi^0$ detector, an
assembly of 768 CsI(Tl) crystals, covers 75\% of the total solid
angle. Since photons produce electromagnetic showers, their energy was
shared among several crystals. The photon energy and hit position were
obtained by summing the energy deposits and energy-weighted centroid,
respectively. Timing information from each module was used to identify a
photon cluster and to suppress accidental backgrounds due to beam
particles. The two-photon invariant mass ($M_{\gamma \gamma}$) and the
$\pi^0$ energy and direction were obtained from the photon momentum
vectors.   
\begin{figure}
\begin{center}
 \epsfig{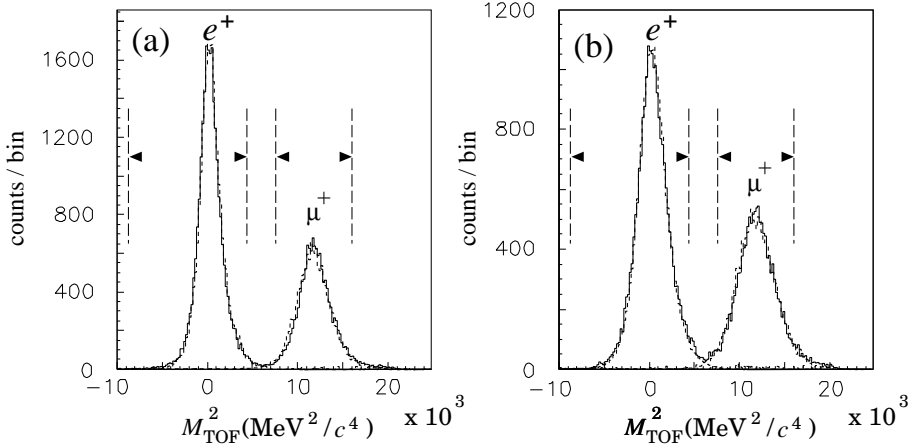}
\end{center}
  \caption{ 
 Mass squared spectra $M^{2}_{\rm TOF}$ obtained in TOF analysis for
 all momenta at (a)0.65 and (b)0.90 T. The solid line is the
 experimental data and the dotted line is the Monte Carlo
 simulation. The Monte Carlo data are scaled by the obtained
 $\Gamma(K_{\mu3})/\Gamma(K_{e3})$ value. The $\mu^+$ and $e^+$
 selection regions are indicated by the dashed lines. 
  }
  \label{fg:massgmc}
\end{figure}

$K_{\mu3}$ and $K_{e3}$ decays at rest were selected by the following
procedure, which is similar to our $K_{e3}$ study.
The $K^+$ decay time, defined as the charged lepton signal at the TOF1
counter, was required to be more than 5 ns later than the $K^+$ arrival time
measured by the ${\rm \check{C}}$erenkov counter to remove in-flight
$K^+$ decays. Events with $\pi^+$ decays in-flight and scattering of
charged particles from the magnet pole faces were eliminated by a track
consistency cut in the ring counters. Events with two clusters in the
CsI(Tl) calorimeter were selected as $\pi^0$ decays and events with
other cluster numbers were rejected. The acceptance cut on the invariant
mass was 50$<M_{\gamma \gamma}<$140 MeV/{\it c}$^2$. Requirements for
the charged particle momentum corrected for the energy loss in the
target ($P_{\rm cor}<$ 175 MeV/$c$) plus opening angle between a charged
particle and the $\pi^0$ ($\theta_{l^+ \pi^0}<$ 154$^{\circ}$) removed
the $K_{\pi 2}$ events. 

\begin{figure}
\begin{center}
  \epsfig{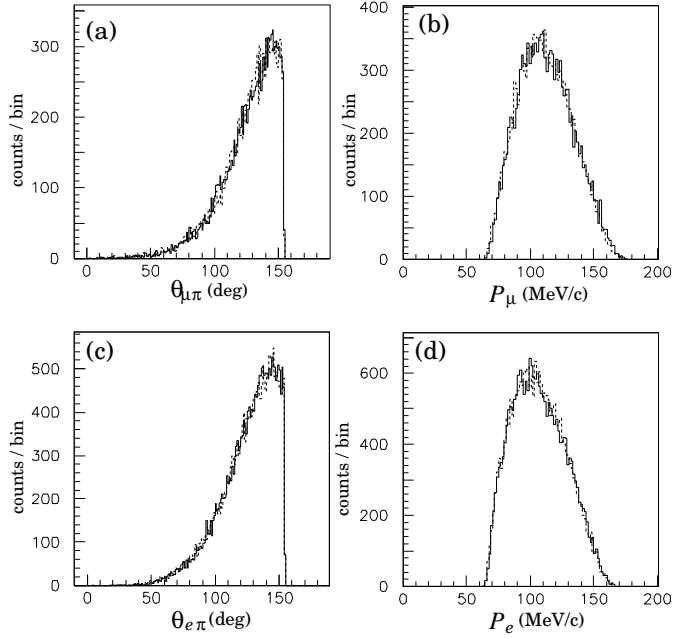}
\end{center}
\caption{Spectra of the selected events (solid line): (a)$\theta_{\mu^+ \pi ^0}$, (b)$P_{\mu^+}$ for $K_{\mu3}$ decay and (c)$\theta_{e^+ \pi ^0}$, (d)$P_{e^+}$ for $K_{e3}$ decay, and Monte Carlo simulation (dotted line).
They were obtained for the setting of B=0.65 T.}
\label{fg:65kl3}
\end{figure}

\begin{figure}
\begin{center}
  \epsfig{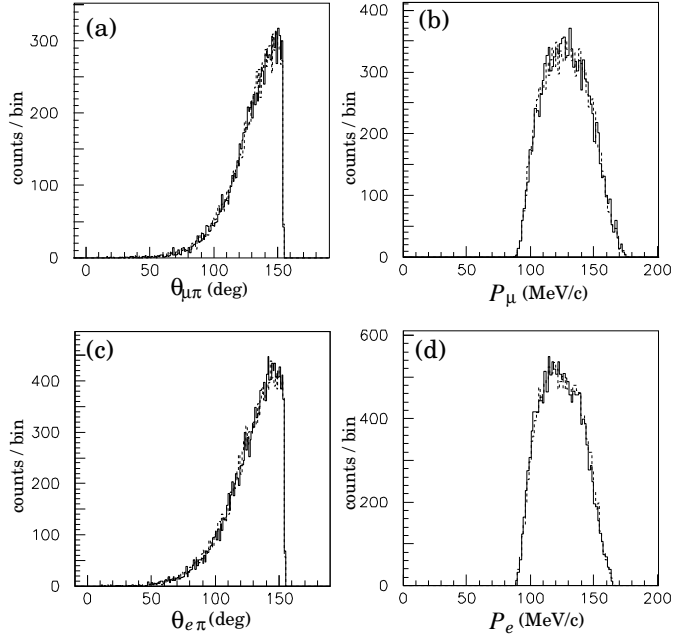}
\end{center}
\caption{Spectra of the selected events (solid line): (a)$\theta_{\mu^+ \pi ^0}$, (b)$P_{\mu^+}$ for $K_{\mu3}$ decay and (c)$\theta_{e^+ \pi ^0}$, (d)$P_{e^+}$ for $K_{e3}$ decay, and Monte Carlo simulation (dotted line).
They were obtained for the setting of B=0.90 T.}
\label{fg:90kl3}
\end{figure}

The mass squared ($M^2_{\rm TOF}$) of the charged particles, obtained
from the TOF and momentum, are shown in Fig.~\ref{fg:massgmc} integrated
over the entire momentum region. The $\mu^+$ and $e^+$ selection regions
are also indicated in the figure. The timing resolution of
$\sigma_T=$270 ps provides a good separation between the $K_{\mu 3}$ and
$K_{e 3}$ events. It should be emphasized that only these requirements
were imposed for the $K_{\mu3}$ and $K_{e 3}$ event extraction. The
numbers of good events after those cuts mentioned above are 12882(10704)
and 23122(16850) at 0.65(0.90) T for $K_{\mu 3}$ and $K_{e3}$ decays,
respectively, where small fractions due to the background events are
still included. Fig.~\ref{fg:65kl3} and~\ref{fg:90kl3} show the $K_{\mu
3}$ and  $K_{e3}$ spectra for the setting of B=0.65 and 0.90 T,
respectively. 
\section{Monte Carlo simulation}
In order to obtain the detector acceptance and estimate the background
fraction, the Monte Carlo simulation was carried out for both the 
charged particle measurement by the spectrometer and the $\pi^0$
measurement by the CsI(Tl) detector. The initial Dalitz distributions
were generated with the values of $\lambda_{+}=0.0278$ and the current
world average $\lambda_{0}=0.006$, while the Dalitz distribution of
$K_{e3}$ decay is insensitive to the $\lambda_{0}$ parameter. The
radiative corrections were taken into account by following the Ginsberg
procedure~\cite{gin70}. The simulation data were analyzed in the same
manner as the experimental sample.  The dotted lines in
Fig.~\ref{fg:65kl3} and Fig.~\ref{fg:90kl3} show the simulation spectra
for the setting of B=0.65 and 0.90 T, respectively, using the assumed
$\lambda_+$ and $\lambda_0$ parameters. These spectra are
normalized so that the total number of events is the same as the
experimental one. Here, it is to be noted that much higher statistical
accuracy for the Monte Carlo simulation is necessary to determine the
form factors from the $K_{\mu3}$ Dalitz plot analysis.

\begin{figure}
\begin{center}
\epsfig{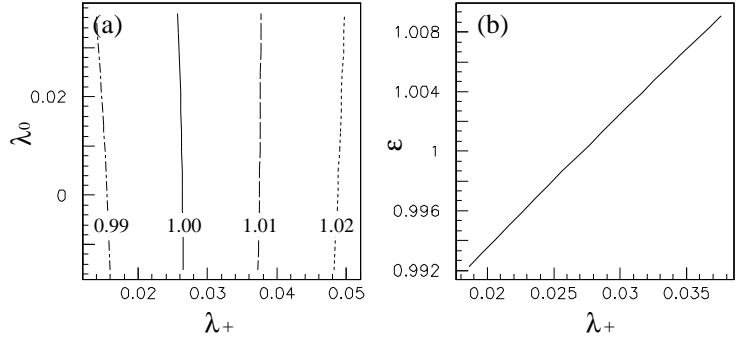}
\end{center}
\caption{
 The correction factor for the detector acceptance for the setting of
 B=0.65 T; (a) $\epsilon$
 contour in terms of the $\lambda_+$ and $\lambda_0$ parameters for
 $K_{\mu3}$ and (b) dependence on the $\lambda_+$ parameter for $K_{e3}$.
The correction factor for the setting of B=0.90 T is similar.
}
\label{fg:lacc}
\end{figure}

By using the simulation data, the detector acceptance ($\Omega$) is
calculated as, 
\begin{eqnarray}
 \Omega = N_{\rm acc}/N_{\rm gen}\times 
\epsilon(\lambda_+,\lambda_0) \equiv \Omega^0 \times 
\epsilon(\lambda_+,\lambda_0), \nonumber
\end{eqnarray}
where $N_{\rm acc}$ and $N_{\rm gen}$ are the number of events accepted
by our event selection requirements and the number of generated $K^+$
events, respectively. $\Omega^0$ denotes the detector acceptance at the
assumed form factor for the production of the simulation samples. They
were $\Omega^0(K_{\mu3})= (1.790\pm0.014)\times 10^{-3}$,
$\Omega^0(K_{e3})= (2.161\pm0.014)\times 10^{-3}$ for the setting of
B=0.65 T and $\Omega^0(K_{\mu3})= (1.669\pm0.014)\times 10^{-3}$,
$\Omega^0(K_{e3})= (1.756\pm0.013)\times 10^{-3}$ for the setting of
B=0.90 T. The detector acceptance depends only slightly on the form
factor ($i.e.$, the shape of the Dalitz distribution), therefore a
correction factor $\epsilon(\lambda_+,\lambda_0)$ is introduced. 
$\epsilon$ was calculated by taking into account the event selection
requirements such as $P_{\rm cor}$, $\theta_{l^+ \pi^0}$, and $M_{\rm
TOF}^2$ cuts. As shown in Fig.~\ref{fg:lacc}(a,b), the $\epsilon$
distribution for $K_{\mu3}$ decay was obtained as a contour plot in the
($\lambda_{+}$,$\lambda_{0}$) space, while $\epsilon$ for $K_{e3}$ decay
was obtained as a function of the $\lambda_+$ parameter.

$M^2_{\rm TOF}$ spectra were calculated by assuming a TOF resolution
of $\sigma_{T}$=270 ps. They are shown as dotted lines in
Fig.~\ref{fg:massgmc}. Since the charged particles were identified by
$M^2_{\rm TOF}$ measurement, the choice of the TOF response function was
important for the determination of the background contamination. We took
into account $K_{\pi 2}$, $K_{e3}$, $K_{\mu3 \gamma}$ decays for the
$K_{\mu 3}$ background and $K_{\pi 2}$, $K_{\mu 3}$, $K_{e3 \gamma}$
decays for the $K_{e3}$ background. The background fraction depends on
the magnetic setting of the spectrometer, as summarized in
Table~\ref{tb:bgke3}. The most dominant background is due to $K_{\pi 2}$
decay in-flight. The accidental background
fraction due to beam particles was estimated to be 0.3\% for both
$K_{\mu3}$ and $K_{e3}$ samples. 
\begin{table}
\begin{center}
\caption{Background fractions included in the $K_{\mu3}$ and $K_{e3}$
 samples. 
}
\label{tb:bgke3}
\begin{tabular}{c|l|rr}
\hline
&background&\multicolumn{2}{c}{background fraction}\\
&item&\multicolumn{1}{c}{B=0.65 T}&\multicolumn{1}{c}{B=0.90 T}\\ \hline 
&$K_{\pi2}$&0.6\%&0.9\% \\ 
$K_{\mu3}$&$K_{e3}$&0.1\%&0.1\% \\ 
&$K_{\mu3 \gamma}$&$<0.1$\%&$<0.1$\% \\ 
&accidental&0.3\%&0.3\% \\ 
\hline
&$K_{\pi2}$&0.4\%&0.3\% \\ 
$K_{e3}$&\multicolumn{1}{|l|}{$K_{\mu3}$}&$<0.1$\%&$<0.1$\% \\ 
&$K_{e3 \gamma}$&$<0.1$\%&$<0.1$\% \\ 
&accidental&0.3\%&0.3\% \\ 
\hline
\end{tabular}
\end{center}
\end{table}

\section{Results}
\begin{figure}
\begin{center}
\epsfig{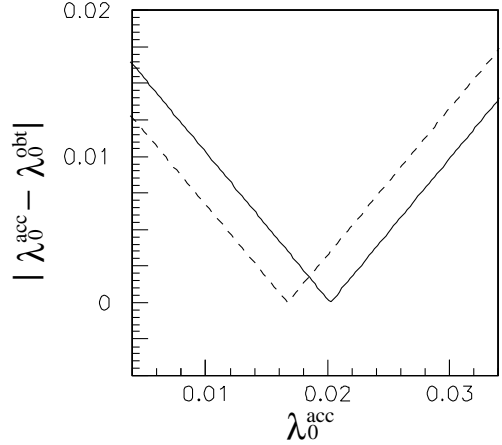}
\end{center}
 \caption{ $| \lambda_0^{\rm acc} - \lambda_0^{\rm obt}|$ plot
 as a function of $\lambda_0^{\rm acc}$. The solid and dashed lines
 correspond to 0.65 and 0.90 T data, respectively. The value of
 $\lambda_0^{\rm acc}$, which satisfied the condition of $| \lambda_0^{\rm acc} - \lambda_0^{\rm obt}|=0$, was adopted as the final result.}   
\label{fg:lam}
\end{figure}

The $\Gamma(K_{\mu 3})/\Gamma(K_{e 3})$ ratio can be written as, 
\begin{eqnarray}
\Gamma(K_{\mu 3})/\Gamma(K_{e 3})= N(K_{\mu 3})/N(K_{e 3})\cdot
 \Omega(K_{e3})/\Omega(K_{\mu 3}), \nonumber
\end{eqnarray}
where $N$ is the number of accepted events after subtracting
backgrounds. The $\lambda_0$ parameter can be determined by substituting
the obtained $\Gamma(K_{\mu 3})/\Gamma(K_{e 3})$ value and
$\lambda_+=0.0278$ into Eq.~(\ref{kihon1}). However, the situation was a
little more complicated because the $K_{\mu 3}$ acceptance also depends 
on the $\lambda_0$ parameter. Therefore, it was derived by iteration to
minimize the difference between the $\lambda_0$ parameter for the
acceptance determination ($\lambda_0^{\rm acc}$) and the $\lambda_0$
parameter obtained from Eq.~(\ref{kihon1}) ($\lambda_0^{\rm
obt}$). Fig.~\ref{fg:lam} shows the $|\lambda_0^{\rm acc} -
\lambda_0^{\rm obt}|$ plot as a function of $\lambda_0^{\rm
acc}$ for both the 0.65 and 0.90 T data. The value of $\lambda_0^{\rm
acc}$, which satisfied the condition of $|\lambda_0^{\rm acc} -
\lambda_0^{\rm obt}|$=0, was adopted as our final result. The $\lambda_0$
parameter and associated $\Gamma(K_{\mu 3})/\Gamma(K_{e 3})$ value are
shown in Table~\ref{tb:results} together with those of the world average
quoted in PDG. The results of B =0.65 and 0.90 T data were then combined
by calculating the error weighted average as, $\lambda_0=0.019\pm0.005$
and $\Gamma(K_{\mu 3})/\Gamma(K_{e3})=0.671\pm0.007$.  
\begin{table}
\begin{center}
\caption{Results of 
$\Gamma(K_{\mu3})/\Gamma(K_{e 3})$ and $\lambda_0$ parameter
}
\label{tb:results}
\begin{tabular}{lcc}
\hline
& $\Gamma(K_{\mu3})/\Gamma(K_{e 3})$ & $\lambda_0$\\
\hline
0.65T & 0.673$\pm$ 0.010 & 0.020$\pm$ 0.006 \\
0.90T & 0.668$\pm$ 0.011 & 0.017$\pm$ 0.007 \\
combined &0.671$\pm$ 0.007 & 0.019$\pm$ 0.005 \\
no $\mu$-$e$ universality assumption&0.669$\pm$ 0.007 & $-$ \\
\hline
world average~\cite{par00} &0.680$\pm$ 0.013 & 0.006$\pm$ 0.007 \\
\hline
\end{tabular}
\end{center}
\end{table}

The systematic errors, which have been categorized into background
contamination and inaccurate detector acceptance, are summarized in
Table~\ref{tb:syserr}. The ambiguity of the $K_{\pi 2}$ fraction 
introduces a significant systematic error, while the other channels
are negligible. If the $K_{\pi 2}$ background in the $K_{\mu3}$ sample is
not correctly evaluated, the results could strongly depend on the cut
points of $\theta_{l \pi^0}$ and $P_{\rm cor}$. The dependence on both
cut points was treated as a systematic error. Also, an imperfect TOF
response function in the simulation could introduce a systematic error,
which would be concentrated around the tail part of the $M_{\rm TOF}^2$
peaks. This contribution was also studied by the cut point dependence of
the TOF window positions. The accidental backgrounds could be neglected
because their fractions were small and common to both the $K_{\mu 3}$
and the $K_{e 3}$ events.  

The systematic errors associated with the measurement, namely
instrumental systematic errors, are negligible. The effects due to
misalignment of the CsI(Tl) barrel, $K^+$ target, and MWPCs and the
contribution of misunderstanding of the energy loss of the charged
particles in the target were estimated and found to be
negligible. Differences in the detection efficiency for $e^+$ and $\mu^+$
in the MWPCs would introduce a systematic error, which was studied by
the sector number dependence of the spectrometer. The results obtained
in the various sectors were distributed within statistical error, and
this contribution was also estimated to be negligible. Since the
detector acceptance depends on the $\lambda_+$ parameter, as well as the
$\lambda_0$ parameter, the ambiguity of the $\lambda_{+}$ parameter
introduced a systematic error. This effect was estimated using the
$\epsilon$ distributions shown in Fig.~\ref{fg:lacc}. Using
$\lambda_+=0.0278\pm0.0040$, the change of the
$\epsilon(K_{\mu3})/\epsilon(K_{e3})$ value in this region was
considered to be the systematic error. 
If the current world average $\lambda_+=0.031\pm0.008$ for $K_{\mu 3}$
decay~\cite{par00} is used instead of taking the $\mu$-$e$
universality value of 0.0278, 
$\Gamma(K_{\mu3})/\Gamma(K_{e 3})$ is shifted to 0.669$\pm$0.007 through
the modification of the $K_{\mu 3}$ acceptance. Although this shift
($-$0.002) is much smaller than the statistical error (0.007), the
acceptance deformation due to the ambiguity of the $K_{\mu 3}$
$\lambda_+$ was included as an additional systematic error. These errors,
regarding them as one standard deviation errors, are summarized in
Table~\ref{tb:syserr}. The total size of the systematic errors was
obtained by
adding each item in quadrature. The total systematic errors for 
$\Delta[\Gamma(K_{\mu3})/\Gamma(K_{e 3})]$ and $\Delta \lambda_0$ are 0.008 and
0.004, respectively, which are basically equal to the statistical error.

\begin{table}
\begin{center}
\caption{Systematic errors}
\label{tb:syserr}
\begin{tabular}{lrr}
\hline
&\multicolumn{1}{c}{$\Delta [\Gamma(K_{\mu3})/\Gamma(K_{e 3})]$}
&\multicolumn{1}{c}{$\Delta \lambda_0$} \\
\hline
$[$Background contamination]&&\\
decay-in-flight from $K_{\pi 2}$ decay& 0.004& 0.003 \\
TOF response&0.004&0.003\\
$[$Detector acceptance]&&\\
instrumental misalignment & $<$0.001& $<$0.001 \\
energy loss estimation in the target & $<$0.001& $<$0.001 \\
experimental error of the $K_{e3}$ $\lambda_+$ parameter & 0.001& 0.001 \\
choice of the $K_{\mu3}$ $\lambda_+$ parameter & 0.005& 
\multicolumn{1}{c}{$-$} \\
\hline
total &0.008&0.004\\
\hline
\end{tabular}
\end{center}
\end{table}
\section{Conclusion}
The ratio of the $K^{+}\rightarrow \pi^{0} \mu^{+} \nu$ ($K_{\mu3}^+$) and
$K^{+}\rightarrow \pi^{0} e^{+} \nu$ ($K_{e3}^+$) decay 
widths, $\Gamma(K_{\mu 3})/\Gamma(K_{e3})$, has been measured for 
stopped positive kaons. Assuming $\mu$-$e$ universality in $K_{l3}^+$
decay, the coefficient of the $q^2$ dependent term of the $f_0$ form factor
was determined from the measured $\Gamma(K_{\mu 3})/\Gamma(K_{e3})$ ratio. 
In contrast to the previous experiments, a large detector acceptance and
its symmetrical structure enabled us to reduce the
statistical errors while suppressing the systematic
errors. Our results are \\
\begin{eqnarray}
\Gamma(K_{\mu3})/\Gamma(K_{e3})&=& 0.671\pm0.007({\rm stat.})\pm0.008({\rm syst.}), \nonumber \\  
\lambda_0&=&0.019\pm0.005({\rm stat.})\pm0.004({\rm syst.}). \nonumber 
\end{eqnarray}

The $\lambda_0$ parameter obtained from the present work is consistent
with that from recent $K_{\mu 3}^+$ Dalitz plot
analyses~\cite{whi80,art97}, which supports the validity of $\mu$-$e$
universality in $K_{l3}^+$ decay. If the assumption of $\mu$-$e$
universality is removed, $\Gamma(K_{\mu3})/\Gamma(K_{e3})$ can be
written as~\cite{fea70},
\begin{eqnarray}
\Gamma(K_{\mu3})/\Gamma(K_{e3})&=&[g_{\mu}f_{+}^{\mu}(0)/g_{e}f_{+}^e(0)]^2 \nonumber \\ 
&\times&(0.6457+2.2342\lambda_+^{\mu}-2.3873\lambda_+^{e}+1.5646\lambda_0), \label{eqnot} 
\end{eqnarray}
where $g$ is weak coupling constant for the lepton current. Substituting\\
$\Gamma(K_{\mu3})/\Gamma(K_{e3})=0.669\pm0.011$ (present result with
no $\mu$-$e$ universality assumption),
$\lambda_+^{\mu}=0.031\pm0.008$ (the current world average), 
$\lambda_+^e=0.0278\pm0.0040$ (our $K_{e3}$ result), and
$\lambda_0=0.039\pm0.011$ (error weighted average of Ref.~\cite{whi80}
and~\cite{art97}) into Eq.~(\ref{eqnot}), $g_{\mu}f_{+}^{\mu
}(0)/g_{e}f_{+}^{e}(0)$ is derived to be $0.971\pm0.019$ which is
consistent with unity within the experimental error. 
From the viewpoint of the theoretical framework, the
$K_{l3}^+$ form factors are of importance in studying low energy
properties of the strong interaction in terms of effective theories. The
predicted values of the $\lambda_0$ parameter are $
\lambda_0=0.017\pm0.004$~\cite{gas85} and $
\lambda_0=-0.03$~\cite{kaz76}. The present result is consistent with the
former and
inconsistent with the latter. Also, the $\Delta I=$1/2 rule leads to
identical $\lambda_0$ parameters between $K_{l3}^+$ and $K_{l3}^0$
decays~\cite{aue67}. The $\lambda_0$ parameter from the $K_{l3}^0$
analyses has been determined to be
$\lambda_0=0.025\pm0.006$~\cite{par00}, which is consistent with the
present result.  

\begin{ack}
This work has been supported in Japan by a Grant-in-Aid from the Ministry
of Education, Science, Sports and Culture, and by JSPS; in Russia by
the Ministry of Science and Technology, and by the Russian Foundation
for Basic Research; in Canada by NSERC and IPP, and by the TRIUMF
infrastructure support provided under its NRC contribution; in Korea by
BSRI-MOE and KOSEF; in the U.S.A by NSF and DOE; and in Taiwan by
NSC. The authors gratefully acknowledge the excellent support
received from the KEK staff.
							    \end{ack}


\begin{thebibliography}{00}
\bibitem{gas85} J. Gasser and H. Leutwyler, Nucl. Phys. {\bf B250}, 517
	(1985).
\bibitem{kaz76} D.I.~Kazakov, V.N.~Pervushin, and M.K.~Volkov, Phys. Lett. {\bf B64}, 201 (1976). 
\bibitem{shi00} S. Shimizu {\em et al.}, Phys. Lett. {\bf B495}, 33
	(2000). 
\bibitem{par00} Particle Data Group, Eur. Phys. J. {\bf C15}, Review of
Particle Physics (2000). 
\bibitem{fea70}  H.W. Fearing, E.~Fischbach, and J.~Smith,
	Phys. Rev. {\bf D2}, 542 (1970); Phys. Rev. Lett. {\bf 24}, 189 
	(1970).
\bibitem{ima:90} J. Imazato {\em et al.}, in {\it Proceedings of the
	11th International Conference on Magnet Technology} (Elsevier
	Applied Science, London, 1990), p.366. 
\bibitem{main:99} M. Abe {\em et al.}, Phys. Rev. Lett. {\bf 83}, 4253
	(1999). 
\bibitem{heintze} J. Heintze {\em et al.}, Phys. Lett. {\bf B70}, 482
	(1977). 
\bibitem{nimcsi} D.V. Dementyev {\em et al.}, Nucl. Instr. Method {\bf
	A440}, 151 (2000).
\bibitem{ring:97} A.P. Ivashkin {\em et al.}, Nucl. Instr. Method {\bf
	A394}, 321 (1997).  
\bibitem{gin70} E.S. Ginsberg, Phys. Rev. {\bf 42}, 1035 (1966).
\bibitem{aue67} L.B. Auerbach {\em et al.}, Phys. Rev. {\bf 155}, 1505
	(1967). 
\bibitem{whi80} R. Whitman {\em et al.}, Phys. Rev. {\bf D21}, 652
	(1980).
\bibitem{art97} V.M. Artemov {\em et al.}, Physics of Atomic and
	Nuclei. {\bf 60} 2023 (1997)
\end{thebibliography}
\end{document}